\documentclass[aps,prl,reprint,amsmath,amssymb,groupedaddress,showpacs,nobibnotes]{revtex4-1}
\usepackage{graphicx}
\usepackage{bm}     % bold math
\usepackage{hyperref} % add hypertext capabilities
\usepackage{dsfont}    % allows \mathds{}
\usepackage{color,SIunits}

\begin{document}

\title{Conditions for two-photon interference with coherent pulses}
%Observation of two-photon interference with temporally non-overlapping coherent pulses}

\author{Yong-Su Kim}
\email{yskim25@gmail.com}
\affiliation{Information Technology Laboratory, National Institute of Standards and Technology, 100 Bureau Dr., Gaithersburg, MD 20899, USA}

\author{Oliver Slattery}
\affiliation{Information Technology Laboratory, National Institute of Standards and Technology, 100 Bureau Dr., Gaithersburg, MD 20899, USA}

\author{Paulina S. Kuo}
\affiliation{Information Technology Laboratory, National Institute of Standards and Technology, 100 Bureau Dr., Gaithersburg, MD 20899, USA}

\author{Xiao Tang}
\email{xiao.tang@nist.gov}
\affiliation{Information Technology Laboratory, National Institute of Standards and Technology, 100 Bureau Dr., Gaithersburg, MD 20899, USA}

\date{\today}

%%%%%%%%%%%%%%%%%%%%%%%%%%%%%%%%%%
\begin{abstract}
We study the conditions for two-photon classical interference with coherent pulses. We find that the temporal overlap between optical pulses is not required for the interference. However, the coherence within the same inputs is found to be essential for the interference.
%We report experiments on two-photon interference between temporally non-overlapping weak coherent pulses. While the single-photon interference is washed out, the two-photon interference shows a  Hong-Ou-Mandel dip with visibility of $0.50\pm 0.09$, which shows that the two-photon classical interference does not require temporal overlapping between optical pulses.
\end{abstract}
%%%%%%%%%%%%%%%%%%%%%

\pacs{34.80.Pa, 42.25.Hz, 42.50.-p}%Coherence and correlation, Interference, Quantum optics

\maketitle

%%%%%%%%%%%%%%%%%%%%%%%%%%%%%%%%%%%%%%%%%%%%%%%%%%%%%%%%%

%Introduction
%Researches on classical and/or quantum interferences has been leading better understandings of nature. For example, Young's double-slit interference helped the wave properties of light which eventually leads to famous Maxwell equations. The Hong-Ou-Mandel interference has been utilized to understand quantum 

%Interference, in classical physics, is described as the coherent superposition of waves. Since Young showed the double slit interference of light, it has usually been showed with light and often considered as a proof of wave property of light. 

Interference is one of the most interesting phenomena in nature for many physicists. Since the first experimental demonstration of optical interference by Young \cite{young 07}, it has been considered one of the most important notions for understanding optics \cite{hecht}. Classically, it is understood as a coherent superposition of electromagnetic waves, and it explains, in classical terms, many interesting phenomena. For example, one of the outputs of a Mach-Zehnder (MZ) interferometer shows a sinusoidal oscillation with respect to the relative phase difference between two inputs and this phenomenon can be fully explained by classical theory.

Classical physics, however, cannot sometimes completely describe interference. Let us consider a Hong-Ou-Mandel (HOM) interference between two identical photons \cite{hong 87, shih 88, ou 88}. When two identical photons enter into a beamsplitter (BS) at the same time, the coincidences between two detectors at the  outputs of the BS are completely suppressed, a characteristic referred as a HOM dip. The visibility, $V$, of the HOM dip is defined as the relative depth of the dip compared to the non-interfering cases. Using single-photon states, the coincidences can be completely suppressed, so the visibility can reach up to $V=1$. The classical theory of the coherent superposition of electromagnetic waves, however, can only explain a HOM dip with $V\le0.5$ \cite{rarity 05}. Thus, a HOM dip with $V>0.5$ should be considered as a non-classical phenomenon, and therefore a quantum effect described by a superposition of indistinguishable probability amplitudes \cite{feynman 85}. Note that the classical HOM interference has been studied with the application to the investigation of temporal and/or spectral properties of light source \cite{ou 89, miyamoto 93, baba 96, li 97, miyamoto thesis}.

%Note that $V=0.5$ is often referred as a classical/quantum boundary of two-photon interference. Note that this quantum theory fully explains not only the exclusive quantum interference phenomena but also 

In many experimental demonstrations with optical pulses, classical and quantum interference are measured when pulses have temporal overlap at a BS \cite{ou 89, miyamoto 93, baba 96, mandel99, kim10, kim11}. It often leads to a common misconception that classical and/or quantum interference requires the the optical pulses to be temporally overlapping. However, it has been shown that temporal overlapping between optical pulses is not a requirement for quantum interference; Both single- and two-photon quantum interference can occur from temporally non-overlapping single-photon states \cite{pittman 96, kim 99, kim 00, kim 03, kim 05}. In these papers, the authors clearly explain the phenomena in terms of quantum physics and the superposition of probability amplitudes with Feynman diagrams \cite{feynman 85}.

%\textcolor{red}{Since interference are originated from the superposition of waves (or probability amplitudes in quantum physics), if one can somehow achieve the superposition of waves or probability amplitudes without temporal overlapping, the interference is still measurable.}

%\textcolor{red}{It has been shown that both single- and two-photon quantum interference does not require the temporal overlapping between optical pulses. 

%The superposition of probability amplitudes between two temporally separated optical pulses can be well represented by Feynman diagrams so it is easy to understand \cite{feynman 85}. Based on this understanding, it has been shown that both single- and two-photon quantum interference can occur from temporally non-overlapping single-photon states \cite{pittman 96, kim 99, kim 00, kim 03, kim 05}. In their papers, the authors clearly showed that quantum interference does not require temporal overlapping but rather indistinguishable probability amplitudes.}

%Since a concept of the superposition between temporally separated probability amplitudes is more intuitive and well represented by Feynman diagram, it is easier to understand in quantum physics. Based on that, it has been shown that both single- and two-photon quantum interference can  occurr from temporally non-overlapping single-photon states \cite{pittman 96, kim 99, kim 00, kim 03, kim 05}. In their papers, the authors clearly showed that quantum interference does not require temporal overlapping but rather indistinguishable probability amplitudes.

In classical physics, however, the superposition of probability amplitudes and Feynman diagrams are not applicable. Instead, the classical electromagnetic waves superposition theory should be employed to describe the interference. It is easy to think that there would be no interference between two temporally non-overlapping optical pulses because it seems that the electromagnetic waves do not exist without an optical pulse. Counter-intuitively, however, the classical interference does not require the temporal overlap of optical pulses \cite{miyamoto thesis, li 97, deBroglie 68}.

%In classical physics, however, the superposition of the waves of temporally non-overlapping pulses is ambiguous because one can easily think that the electromagnetic waves do not exist when there is not an optical pulse. Despite the ambiguity, there has been research which explore the effects of the temporal bias of the classical interferometer. For instance, the single-photon classical interference with either a mode-locked Ti:Sapphire laser or a multi-mode cw diode laser has been shown to be revived   even when the temporal biasing is much larger than the coherent length of optical pulse under certain circumstances \cite{baek 07}. These results, however, do not satisfy the aforementioned condition. For mode-locked laser pulses, the interference occurs between two adjacent laser pulses, so two photons are, actually, temporally overlapped. For a multi-mode cw diode laser, the overlapping of optical pulses is not well defined. 

In this Letter, we study the conditions for two-photon interference between two classical optical pulses. In particular, we investigate the HOM-type two-photon interference with coherent pulses. We found that classical two-photon interference requires the coherence within each input rather than the temporal overlap of optical pulses from the inputs. The result can be explained by the classical theory of waves superposition. We also provide a quantum analogy to this phenomenon for more intuitive understanding.

%While the single-photon interference is suppressed by the relative phase randomization between inputs, a classical visibility limited HOM dip is measured. \textcolor{red}{It is explained by the classical theory of superposition of electromagnetic waves, and thus shows that the classical interference occurs between temporally non-overlapping optical pulses unlike to the intuition.} We also provide a quantum analogy to this phenomenon for intuitive understanding.

%%%%%%%
\begin{figure}[b]
\centering
\includegraphics[width=3.3in]{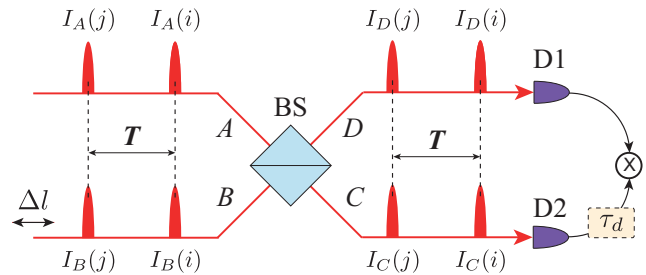}
\caption{The schematic of Hong-Ou-Mandel interference with four weak coherent pulses.}
\label{concept}
\end{figure}
%%%%%%%

%Theory

Figure~\ref{concept} shows the schematic of our two-photon interference experiment with weak coherent pulses, $I_A(i),$ $I_A(j),$ $I_B(i),$ and $I_B(j)$. Here, $i, j$ denote the timing labels and the subscripts $A, B$ are the input modes. The intervals between $I_k(i)$ and $I_k(j)$ for both $k=A,B$ are the same, $T$. The optical delay between two inputs, $\Delta l=I_A(i)-I_B(i)$, can be scanned for the interference measurement. Note that the interval between $I_B(i)$ and $I_B(j)$ is fixed at $T$ during the scanning. We will only consider the case that the scanning of $\Delta l$ is much smaller than $T$, so it does not provide temporal overlap between two different labeling pulses, e.g., between $I_A(i)$ and $I_B(j)$. If the correlation measurement between two outputs $C$ and $D$ is of interest, the electronic delay $\tau_d=0$ or $\pm T$ at $C$ introduces the intensity correlation measurement between pulses at various delays. Note that the electronic delay $\tau_d$ is not interchangeable with the optical delay $\Delta l$. The intensities at the BS outputs $I_C(i)$ and $I_D(i)$ are
%The two sequences of the coherent pulses have the same repetition rates ($1/\textrm{repetition rate}=T$) are sent to a BS at mode $A$ and $B$. Because we have sequences of pulses, we can label each pulse with parameters $i, i+1,\cdots, j$ as depicted in Fig.~\ref{concept}. The optical path delay  between two pulse trains, $\Delta l$, can be scanned for the interference measurement. We will only consider the case that the scanning of $\Delta l$ is much smaller than $T$, so it does not provide temporal overlap between two adjacent pulses, e.g., between $I_A(i)$ and $I_B(i+1)$. The intensities at the BS outputs $I_C(i)$ and $I_D(i)$ are
%%%%%%%
\begin{eqnarray}
I_C(i)&=&\frac{1}{2}I_A(i)+\frac{1}{2}I_B(i)-\sqrt{I_A(i)I_B(i)}\sin{\Delta\phi(i)}\nonumber\\
I_D(i)&=&\frac{1}{2}I_A(i)+\frac{1}{2}I_B(i)+\sqrt{I_A(i)I_B(i)}\sin{\Delta\phi(i)},
\label{Classical HOM1}
\end{eqnarray}
%%%%%%%
where $\Delta\phi(i)$ denotes the relative phase between two pulses $I_A(i)$ and $I_B(i)$. The relative phase can be represented as
%%%%%%%
\begin{equation}
\Delta\phi(i)=\Delta\phi_{AB}(i)+\frac{2\pi\Delta l}{\lambda},
\label{phase}
\end{equation}
%%%%%%%
where $\Delta\phi_{AB}(i)$ represents the inherent phase difference between two pulses $I_A(i)$ and $I_B(i)$ and $\lambda$ is the wavelength of the light.

When two pulses $I_A(i)$ and $I_B(i)$ are coherent, that is $\Delta\phi_{AB}(i)$ has a fixed definite value, Eq.~(\ref{Classical HOM1}) shows sinusoidal interference which corresponds to a single-photon interference, i.e., Mach-Zehnder like interference. However, if the two pulses are incoherent, and thus $\Delta\phi_{AB}(i)$ varies randomly, the single-photon interference will be washed out since $\langle\sin\Delta\phi(i)\rangle=0$, where $\langle x\rangle$ represents the average of $x$ over many events.

The coincidences between D1 and D2 correspond to the correlation measurement between $I_C$ and $I_D$ for low input intensities $I_A$ and $I_B$. Let us first consider the correlation measurement between two pulses at the same timing, $\langle I_C(i)I_D(i)\rangle$. Note that this case is equivalent to a standard HOM interferometer with coherent pulses as the two pulses meet at the BS. It is easily accomplished in the experiment by putting a zero-electronic delay at D2, $\tau_d=0$. Since $\langle\sin\Delta\phi(i)\rangle=0$ for a randomized $\Delta\phi(i)$,  the correlation measurement is represented by \cite{rarity 05}
%%%%%%%
\begin{equation}
\langle I_CI_D\rangle=\frac{1}{4}\langle I_A^2\rangle+\frac{1}{4}\langle I_B^2\rangle+\left(\frac{1}{2}-\langle\sin^2\Delta\phi\rangle\right)\langle I_A\rangle\langle I_B\rangle.
\label{I_C I_D}
\end{equation}
%%%%%%%
Here, we omitted the label $i$ in Eq.~(\ref{I_C I_D}). The $\langle\sin^2\Delta\phi\rangle$ term vanishes for the interference free case. In the interference case, $\langle\sin^2\Delta\phi\rangle=1/2$, thus the whole last term of Eq.~(\ref{I_C I_D}) disappears. Therefore, the visibility of the two-photon interference in classical physics is 
%%%%%%%
\begin{equation}
V_c=\frac{2\langle I_A\rangle\langle I_B\rangle}{\langle I_A^2\rangle+\langle I_B^2\rangle+2\langle I_A\rangle\langle I_B\rangle}.
\label{classical visibility}
\end{equation}
%%%%%%%
For constant intensities $\langle I_k^2\rangle=\langle I_k\rangle^2$, the maximum classical visibility is $V_c^{\textrm{max}}=0.5$ when $\langle I_A\rangle=\langle I_B\rangle$. This result is the classical limit of a HOM interference. 

%Irregular HOM
Now, let us consider the correlation measurement between pulses which did not exist at the same time, i.e., $\langle I_C(i)I_D(j)\rangle$ where $i\neq j$, thus, $\tau_d=T$. Noting that $\langle\sin\Delta\phi(i)\rangle=\langle\sin\Delta\phi(j)\rangle=0$, the correlation measurement can be represented by
%%%%%%%
\begin{widetext}
\begin{eqnarray}
\langle I_C(i)I_D(j)\rangle&=&\frac{1}{4}\langle I_A(i)I_A(j)\rangle+\frac{1}{4}\langle I_A(i)I_B(j)\rangle+\frac{1}{4}\langle I_B(i)I_A(j)\rangle+\frac{1}{4}\langle I_B(i)I_B(j)\rangle\nonumber\\
&-&\sqrt{\langle I_A(i)I_A(j)I_B(i)I_B(j)\rangle}\langle\sin\Delta\phi(i)\sin\Delta\phi(j)\rangle.
\label{different timing I_C I_D}
\end{eqnarray}
\end{widetext}
%%%%%%%
In general, $\langle\sin\Delta\phi(i)\sin\Delta\phi(j)\rangle=0$, thus, Eq.~(\ref{different timing I_C I_D}) does not show interference. However, let us consider the case when intensities of two pulses are the same at the same input, e.g., $I_A(i)$ and $I_A(j)$, and also they are coherent. Then, one can find the following conditions are satisfied:
%%%%%%%
\begin{eqnarray}
I_k(i)&=&I_k(j)\nonumber\\
\Delta\phi(i)&=&\Delta\phi(j)+\Delta\phi_{ij},
\label{long term coherence}
\end{eqnarray}
%%%%%%%
where $\Delta\phi_{ij}$ is a constant phase. Note that the second condition is satisfied if $\Delta\phi_{AB}(i)=\Delta\phi_{AB}(j)+\Delta\phi_{ij}$. It is important to remember that although $\Delta\phi(i)$ and $\Delta\phi(j)$ are related, they are still randomly varying. This condition can be obtained from, for example, mode-locked laser pulses. Note that for the mode-locked laser pulses, $\Delta\phi_{ij}=0$. With these conditions, Eq.~(\ref{different timing I_C I_D}) transforms to Eq.~(\ref{I_C I_D}), and therefore the two-photon interference with a visibility, $V= 0.5$ will be measured.

It is notable that all the events registered as the coincidences for this case come from two temporally separated coherent pulses, one at $i$ and the other at $j$. It seems natural that when there is no overlap between optical pulses, the electromagnetic waves are also not overlapped, i.e., no superposition. This intuition, however, is incorrect as indicated by the last term of Eq.~(\ref{different timing I_C I_D}). This interference term contains all four intensities of input pulses, thus, shows all electromagnetic waves interfere although there is no optical pulse overlapping. In short, in the classical view of interference, the electromagnetic waves are responsible for the interference, not the photons \cite{deBroglie 68}.

%%%%%%
\begin{figure}[t]
\centering
\includegraphics[width=3.3in]{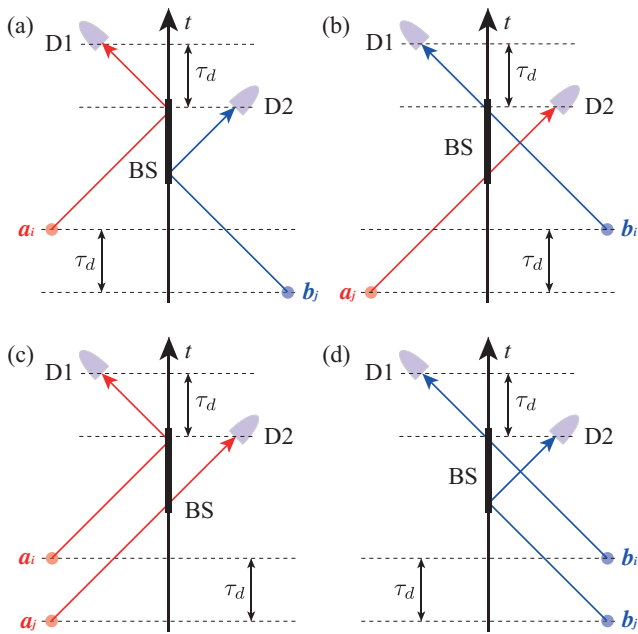}
\caption{Feynman diagrams for $\Delta l=0$. In general, all (a)-(d) cases are distinguishable, so they do not interfere. However, when pulses in the same inputs (between $a_i$ and $a_j$, for example) are coherent, (a) and (b) become indistinguishable, so they interfere. For single-photon pulses, one can effectively suppress (c) and (d) while maintaining the coherence, so one can measure a $V=1$ HOM dip with temporally non-overlapping pulses. For classical optical pulses such as coherent states, however, one cannot remove cases (c) and (d) without disturbing the coherence, so the maximum visibility is limited to 0.5.}
\label{feynman}
\end{figure}
%%%%%%

%Feynman diagram

Because classical physics is a subset of quantum physics, one should be able to explain the interference with temporally non-overlapping coherent pulses with quantum descriptions. Moreover, quantum descriptions are usually more intuitive, so one can more easily understand the physics behind. Therefore, let us consider quantum interpretations of the phenomenon with Feynman diagrams. Since the coincidences are registered by two-photons separated by $T=\tau_d$, there are four possible biphoton amplitudes as depicted in Fig.~\ref{feynman}. Here, $a$ and $b$ denote the annihilation operators at input $A$ and $B$, and the subscripts $i$ and $j$ are the labeling parameters. Although the Feynman diagrams are depicted with single-photon states, it is still applicable for our case since the coherent pulses are so weak that they mostly contain only a single photon and also when more than two-photons exist at the same time, they do not lead to relevant coincidences.

In general, all four cases are distinguishable, so they do not interfere. However, when $a_i$ and $a_j$ are coherent and $b_i$ and $b_j$ are also coherent, Fig.~\ref{feynman} (a) and (b) become indistinguishable, so they do interfere. Because half of the cases interfere while the other half does not, the expected visibility will be 0.5.

This quantum description raises an interesting question: If coherent pulses only exist in the cases shown in Fig.~\ref{feynman} (a) and (b), do they show $V=1$ two-photon interference? The answer to this question is that it is impossible to remove the other cases (Fig.~\ref{feynman} (c) and (d)) while maintaining the coherence between pulses in the same inputs unless the optical pulses are single-photon states \cite{mandel 83, ou 88b}. Thus, even with the quantum description, the classical visibility is limited to 0.5.

It is interesting to compare the visibility limitation scenario to that of Franson interference \cite{franson 89}. When the coincidence detection does not distinguish between the `long-short' and `long(short)-long(short)' cases, the visibility is limited 0.5 \cite{kwiat 90, ou 90}. This visibility limitation in the Franson interferometer can be overcome once the measurement apparatus can distinguish them \cite{kwiat 93} while our visibility limitation is inherent.

%%%%%%%%%%%%%%%%%%%%%%%%%%%%%%%%%%%%%%%%%%%%%%%%%%%

% Experiment

\begin{figure}[t]
\centering
\includegraphics[width=3.3in]{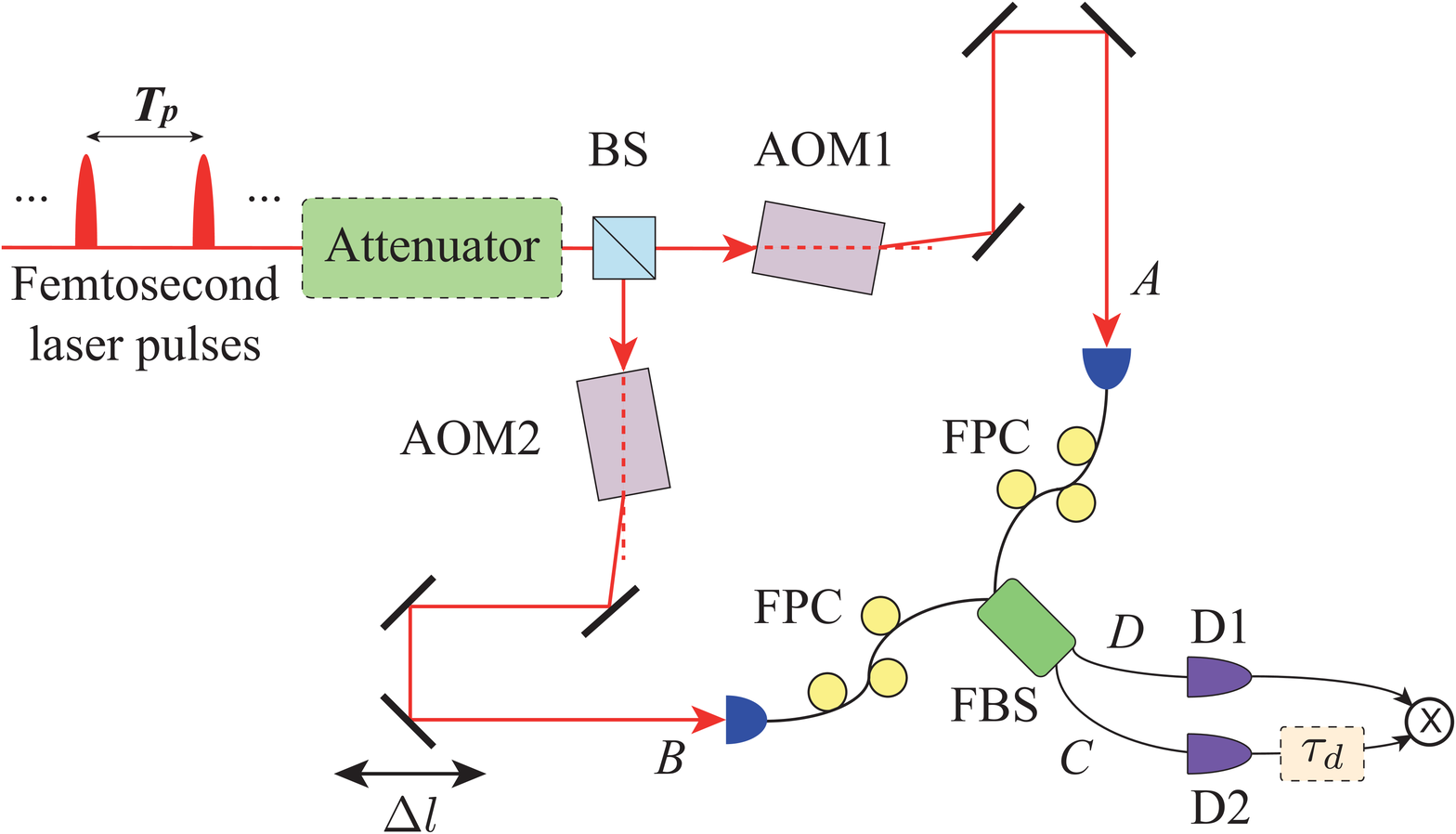}
\caption{Experimental setup. BS: beam splitter, AOM: acousto-optic modulator, FPC: fiber polarization controller, FBS: fiber beamsplitter, D1 and D2: single photon detectors. The AOMs are used for the phase randomization between two arms.}
\label{scheme}
\end{figure}

Figure~\ref{scheme} shows our experimental setup. Femtosecond laser pulses from a mode-locked Ti:Sapphire laser is used for the experiment. Note that the pulse train and the electronic delay $\tau_d=T=mT_p$ where $T_p$ and $m$ are the pulse period and an integer, respectively, will implement Fig.~\ref{concept}. The central wavelength and the spectral bandwidth of the pulses are 780~nm and 15~nm, respectively. The repetition rate of the pulses is 85~MHz, so the interval between adjacent pulses $T_p\approx 11.8~\nano\second$, which corresponds to 3.5~m in space. The attenuator is introduced to reduce the average photon number per pulse. 
%During the experiment, $\bar{n}$ was chosen to \textcolor{red}{0.00}.

%%%%%%%
%%%%%%%
%%%%%%%
%\begin{figure}[b]
%\centering
%\includegraphics[width=3.2in]{4.data1.eps}
%\caption{Single-photon interference when RF signals to AOM1 and 2 are (a) synchronized and (b) independent. The inset of (a) shows a fine scanning near $\Delta l=0$. For synchronized RF signals, a standard Mach-Zehnder interference is measured whereas the independent RF signals erased the interference.}
%\label{data1}
%\end{figure}
%%%%%%%

%%%%%%%
\begin{figure*}[t]
\centering
\includegraphics[width=7in]{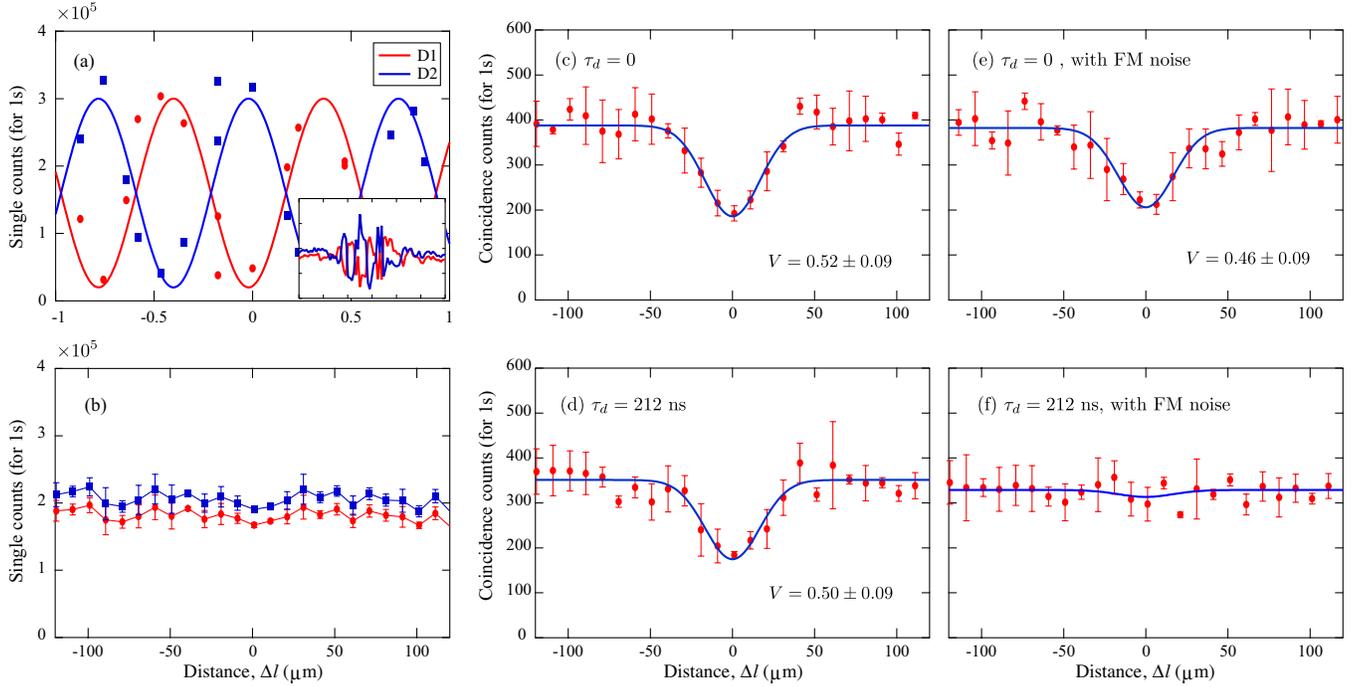}
\caption{Single and coincidence counts for various conditions. Single and coincidence counts are proportional to $\langle I_C\rangle$ ($\langle I_D\rangle$) and $\langle I_CI_D\rangle$, respectively. Error bars are the experimentally obtained standard deviations. The single counts when RF signals to AOM1 and 2 are (a) synchronized and (b) independent. The inset of (a) shows the envelop of the single-photon interference. (c)-(f) coincidence counts between D1 and D2. (c) $\tau_d=0$. (d) $\tau_d=212~\nano\second$. (e), (f) $\tau_d=0$ and $\tau_d=212~\nano\second$ with an additional noise input to the FM input of the RF driver to AOM1. Estimated visibilities are also shown. (c) and (e) correspond to a HOM interference with classical pulses so a standard HOM dip is observed. (d) and (f) correspond to the two-photon interference between temporally non-overlapping optical pulses.}
\label{data}
\end{figure*}
%%%%%%%

A BS splits the incoming pulses into two paths. Each pulse enters into acousto-optical modulators (AOM1 and AOM2) and the deflected pulses are collected by single-mode optical fibers at inputs $A$ and $B$. After the fiber polarization controllers (FPC), that make the polarization identical, the incoming pulses interfere at the fiber beam splitter (FBS). The optical path delay $\Delta l$ is scanned by a translation stage placed at input $B$. A typical scanning range for $\Delta l$ is hundreds of $\micro\meter$ which is much smaller than $T_p$, so the scanning of $\Delta l$ does not provide temporal overlap between adjacent pulses. Silicon avalanche photodiode based single photon detectors D1 and D2 are placed at the outputs of FBS, and a variable electronic delay $\tau_d$ is introduced at D2.

%The whole experimental scheme corresponds to a simple Mach-Zehnder interferometer if the coherence between pulses at different arms are maintained and the single counts at D1 and/or D2 will show sinusoidal oscillations with an large envelop corresponds to the coherence length of the pulse. In order to erase the single-photon interference, one needs to do the phase randomization between these two arms.

The phase randomization between inputs $A$ and $B$ can be accomplished with the help of two AOMs modulated by two independent radio frequency (RF) drivers. An AOM adds additional phase to the deflected beam relative to the driving RF signal. Thus, if the RF signals are unsynchronized, two AOMs will wash out the phase relation between the two inputs. It is experimentally verified by applying either synchronized (see Fig.~\ref{data} (a)) or independent (see Fig.~\ref{data} (b)) RF signals. While the synchronized RF signals maintain the single-photon interference as they conserve the phase relation between $A$ and $B$, the independent RF signals completely suppress the phase interference. Despite the independency of RF signals, the frequencies are still almost the same, 40~MHz. Note that the RF signals do not disturb the coherence within each input, e.g., between $I_A(i)$ and $I_A(j)$.

%Two-photon interference

After we confirmed the phase randomization between $A$ and $B$, we measured  coincidence counts between D1 and D2. The result with $\tau_d=0$ is shown in Fig.~\ref{data} (c). It shows a clear HOM interference with visibility $0.52\pm 0.09$ which is consistent to the classical limit of HOM interference visibility. Note that this case corresponds to a standard HOM interference with temporally overlapped coherent pulses.

Fig.~\ref{data} (d) shows the two-photon interference between temporally non-overlapped  coherent pulses. Here, the electronic delay $\tau_d=212~\nano\second$, so $m=18$ was chosen. The data shows a clear two-photon interference with $V=0.50\pm0.09$. Since the mode-locked laser pulse trains satisfy the coherence condition within the same inputs, Eq.~(\ref{long term coherence}), we can still observe the same two-photon interference as the temporally overlapped pulses.%The mode-locked laser pulse train satisfies the condition of Eq.~(\ref{long term coherence}), and thus shows the same two-photon interference as the temporally overlapped pulses.

%To see how the two-photon interference changes when coherence between pulses at the same inputs is not maintained anymore, so the condition of Eq.~(\ref{long term coherence}) fail. 

In order to disturb the coherence within the same inputs, that is to prevent the pulses satisfying Eq.~(\ref{long term coherence}), we input fast random noise to the frequency modulation (FM) input of one of the AOM RF drivers. The random noise produces random frequency deviations of $\pm50~\%$ to the RF signal, so the coherence between pulses at the same input will be degraded when $\tau_d$ is sufficiently large. This will cause $\Delta\phi_{AB}(i)\neq\Delta\phi_{AB}(j)$, thus $\Delta\phi(i)\neq\Delta\phi(j)$. Note that the amount of the frequency deviation is much smaller than the spectral bandwidth of the coherent pulses, so the interference degradation due to the frequency mismatch is negligible.

The coincidence counts with FM noise input are depicted in Fig.~\ref{data} (e) and (f) for $\tau_d=0$ and 212~ns, respectively. For the case of $\tau_d=0$, the experiment corresponds to a standard HOM interferometer with two temporally overlapping coherent pulses, so we observe the HOM dip with classical visibility limit. The measured visibility is $0.46\pm0.09$. When $\tau_d=212~\nano\second$, however, the two-photon interference is eliminated as the FM  noise diminishes the coherence between pulses at the same input.

To summarize, we studied the conditions for two-photon interference with weak coherent pulses. While the single-photon interference was erased by two AOMs modulated by independent RF signals, two-photon interference with $V=0.5$ was still measured. We found that, counter-intuitively, the {\it classical} two-photon interference requires coherence within each input rather than the temporal overlap of optical pulses from the inputs.

%To summarize, we reported two-photon interference with weak coherent pulses from a single laser. While the single-photon interference was erased by two AOMs modulated by independent RF signals, the two-photon interference with $V=0.5$ was still measured. We observed the same two-photon interference with two temporally separated coherent pulses, thus, clearly verifying that {\it classical} two-photon interference does not require the temporal overlapping of photons. We also  showed that coherence between coherent pulses at the same inputs is essential for the two-photon interference with temporally non-overlapping weak coherent pulses.

%\section{acknowledgments}
The authors thank Y.-W. Cho and Y.-H. Kim for useful discussions. YSK acknowledges the support of the National Research Foundation of Korea (2012R1A6A3A03040505).

%%%%%%%%%%%%%%%%%%%%%%%%%%%%%%%%%%%%%%%%%%%%%%%%%%%

\end{document}